\documentclass[aps,prc,twocolumn,floatfix]{revtex4}

\usepackage{epsfig}

\begin{document}

\title{Differential elliptic flow of identified hadrons and constituent quark number scaling at FAIR}
\author{Partha Pratim Bhaduri}
\affiliation{Variable Energy Cyclotron 
Centre, 1/AF Bidhan Nagar, Kolkata 700 064, India}            
\author{Subhasis Chattopadhyay}
\affiliation{Variable Energy Cyclotron 
Centre, 1/AF Bidhan Nagar, Kolkata 700 064, India}            
\date{\today}

\begin{abstract}
Differential elliptic flow $v_2(p_{T})$ for identified hadrons has been investigated in the FAIR energy regime, employing a hadronic-string transport model (UrQMD) as well as a partonic transport model (AMPT). It has been observed that both the models show a mass ordering of $v_2$ at low $p_{T}$ and a switch over resulting a  baryon-meson crossing at intermediate $p_{T}$. AMPT generates higher $v_2$ values compared to UrQMD. In addition, constituent quark number scaling behavior of elliptic flow has also been addressed. Scaling behavior in terms of the transverse momentum $p_T$ is found to absent for both the partonic as well as the hadronic model. However UrQMD and AMPT with string melting scenario do exhibit a NCQ scaling of $v_2$ at varying degree, with respect to transverse kinetic energy $KE_T$. But default AMPT, where partonic scatterings are not included, does not show any considerable scaling behavior. A variable $\alpha$ is defined to quantify the degree of $KE_T$ scaling. We found that UrQMD gives better scaling than AMPT at FAIR.
\end{abstract}

\maketitle
\section{Introduction}

Statistical QCD predicts that at high temperature and/or density, hadronic matter undergoes a phase transition to a new state, where strongly interacting matter shows partonic behavior. It is generally believed that in the laboratory the only way to produce such a novel state of matter is to collide two heavy nuclei at ultra-relativistic energies. In the past two decades, relativistic heavy ion collision experiments  are performed in different parts of the world with the ultimate aim to map the QCD phase diagram and to discover this new phase of strongly interacting matter, the so-called Quark Gluon Plasma (QGP), where the sub-nuclear degrees of freedom come into play over the nuclear volume rather than the nucleonic volume. In heavy-ion experiments at CERN-SPS and BNL-RHIC, the QCD phase diagram is studied in the region towards high temperatures and low net baryon densities. In future heavy-ion experiments at the CERN-LHC the research program will be continued in the direction of  higher temperature and smaller net baryon densities. The QCD phase diagram is much less explored in the region of high baryon densities and moderate temperatures. In the upcoming Compressed Baryonic Matter (CBM) experiment~\cite{sub}, at the future accelerator facility FAIR at GSI, heavy ions will be collided in the beam energy range between 10 and 40 AGeV. The highest densities (6 to 12 times the normal nuclear matter density) are expected to be produced at the center of the collision zone~\cite{density}.

Since the idea that relativistic nuclear collisions can lead to the formation of a quark-gluon phase has been perceived, lots of theoretical as well as experimental efforts have been devoted to predict unambiguous and experimentally viable probes to indicate the production of the dense partonic medium. The collective flow of the produced particles in the transverse plane of the collision has been long predicted as a signature of the creation of a hot and dense secondary medium~\cite{Greiner,Voloshin}. Of particular interest is the elliptic flow parameter $v_2$, signaling a strong evidence for the creation of a hot and dense system at a very early stage in non-central collisions~\cite{Ollitrault,Voloshin,v2_sorensen}. In RHIC experiments at BNL, a large elliptic flow has been observed~\cite{v2_exp}, as large as predicted by ideal hydrodynamical models~\cite{v2_hydro}. The most striking observation, in this respect, is the number of constituent quark (NCQ) scaling of $v_2$ of identified hadrons.

In the present work, we have explored the $p_T$ dependence of the elliptic flow parameter $v_2$ at top ($E_{Lab} = $ 40 AGeV) and intermediate ($E_{Lab} = $ 25 AGeV) FAIR energies for different types of strange and non-strange hadrons. For this purpose we have employed two transport models: Ultra-relativistic Quantum Molecular Dynamics (UrQMD)~\cite{urqmd} and A Multi-Phase Transport Model (AMPT)~\cite{ampt}. Apart from the differential elliptic flow ($v_2(p_T)$), we have also investigated the constituent quark number scaling behavior of $v_2$ predicted by these models. Finally, we have defined a parameter $\alpha$ to quantify the degree of scaling produced by these two models.

The paper is organized in the following way. In the section II we have presented the results for $p_T$ dependence of $v_2$ for identified hadrons, as predicted by the transport models UrQMD and AMPT. The results from both the models have been compared and contrasted with the existing data from NA49 experiment at 40 AGeV. In the section III we have addressed the issue of  the NCQ scaling of $v_2$ of the selected hadrons. Finally in section IV we summarize our results and conclude.

\section{Differential elliptic flow ($v_2$ vs. $p_T$)}

In the non-central nuclear collisions, the extended overlap area is roughly almond like in shape and does not posses azimuthal symmetry. The initial asymmetries in the geometry of the system give rise to anisotropic pressure gradients which in turn can lead to the anisotropies of the particle momentum distributions. Since the spatial asymmetries decrease rapidly with time, anisotropic flow can develop at very early time of evolution. In that way, the properties of the hot dense matter formed during the initial stage of heavy ion collisions can be learned by measuring the anisotropic flow. The anisotropic flow is defined as the $n$th Fourier coefficient $v_{n}$ of the particle distributions in emission azimuthal angle with respect to the reaction plane~\cite{Voloshin, v2_sorensen}, which can be written as
\begin{equation} 
\frac{dN}{d\phi} \propto
1+2\sum v_n\cos(n(\phi - \Psi)), 
\end{equation}
where $\phi$ denotes the azimuth angle  of the particle, and $\Psi$ denotes the orientation of the reaction plane, defined as the plane spanned by the impact parameter vector and incoming beam direction. The second Fourier coefficient $v_{2}$ represents the elliptic flow which characterizes the eccentricity of the particle distributions in momentum space. In a given rapidity window, it reads as 
\begin{equation}
v_{2}=\langle\cos(2(\phi - \Psi))\rangle=\langle\frac{p_x^2-p_y^2}{p_T^2}\rangle.
\end{equation}

In this paper, we have estimated $v_2$ as  a function of transverse momentum $p_T$ for mid-central (b = 5-9 fm.) Au+Au collisions, at top ($E_{Lab} =$ 40 AGeV) and intermediate ($E_{Lab} =$ 25 AGeV) FAIR energies, in the mid-rapidity region (-1 $ \le y_{cm} \le$ 1). In the present model calculations, the  reaction plane angle is taken as zero so $v_2$ is calculated directly from its definition. 

UrQMD is a hadronic-string transport model based on the same principles as Quantum Molecular Dynamics (QMD)~\cite{QMD} and Relativistic Quantum Molecular Dynamics (RQMD)~\cite{RQMD}. It incorporates a vastly extended collision term with full baryon-antibaryon symmetry. Isospin is explicitly treated for all hadrons. Cross sections in UrQMD, depend on the particle types, their isospins and their centre of mass energy. Within the framework of the UrQMD model, a typical heavy-ion collision proceeds in schematically three stages: the pre-hadronic stage described in terms of strings and constituent di-quarks, the hadronic pre-equilibrium stage and finally the evolution towards hadronic kinetic equilibrium and freeze-out. Particle production in UrQMD either takes place via the decay of a meson or baryon resonance or  via  a string excitation and fragmentation. In this model, partonic interactions are not included except from quark coalescence during the string break-up.  

On the other hand, AMPT is a hybrid transport model, which models an ultra-relativistic nuclear collision incorporating many tools of Monte Carlo simulation. As the initial condition it uses minijet partons from hard processes and strings from soft processes in the Heavy Ion Jet Interaction Generator (HIJING) model~\cite{Hijing}. Time evolution of resulting minijet partons is then described by Zhang's parton cascade (ZPC) model~\cite{ZPC} which includes only parton-parton elastic scatterings with an in-medium cross section derived from pQCD with an effective gluon screening mass taken as a parameter for fixing the magnitude and angular distribution of parton scattering cross section. The subsequent transition from the partonic matter to the hadronic matter is based on the Lund string fragmentation model as implemented in the PYTHIA~\cite{Pythia} generator, where the minijet partons, after they stop interacting, are combined with their parent strings, as in the HIJING model with jet quenching, to fragment into hadrons. The final-state hadronic scatterings are modeled by the ART model~\cite{ART}. In the default AMPT model, minijets coexist with the remaining part of their parent nucleons, and together they form new excited strings, then the resulting strings fragment into hadrons following the Lund string fragmentation. However in the string melting scenario, these strings are converted to soft partons. Interactions among these partons are again described by the ZPC parton cascade model. Since there are no inelastic scatterings, only quarks and antiquarks from the melted strings are present in the partonic matter. The transition from the partonic matter to hadronic matter is then achieved using a simple coalescence model, which combines  two nearest quark and antiquark into mesons and three nearest quarks or antiquarks into baryons or anti-baryons that are close to the invariant mass of these partons.

\begin{figure*} \vspace{-0.1truein}
\includegraphics[width=16.0cm]{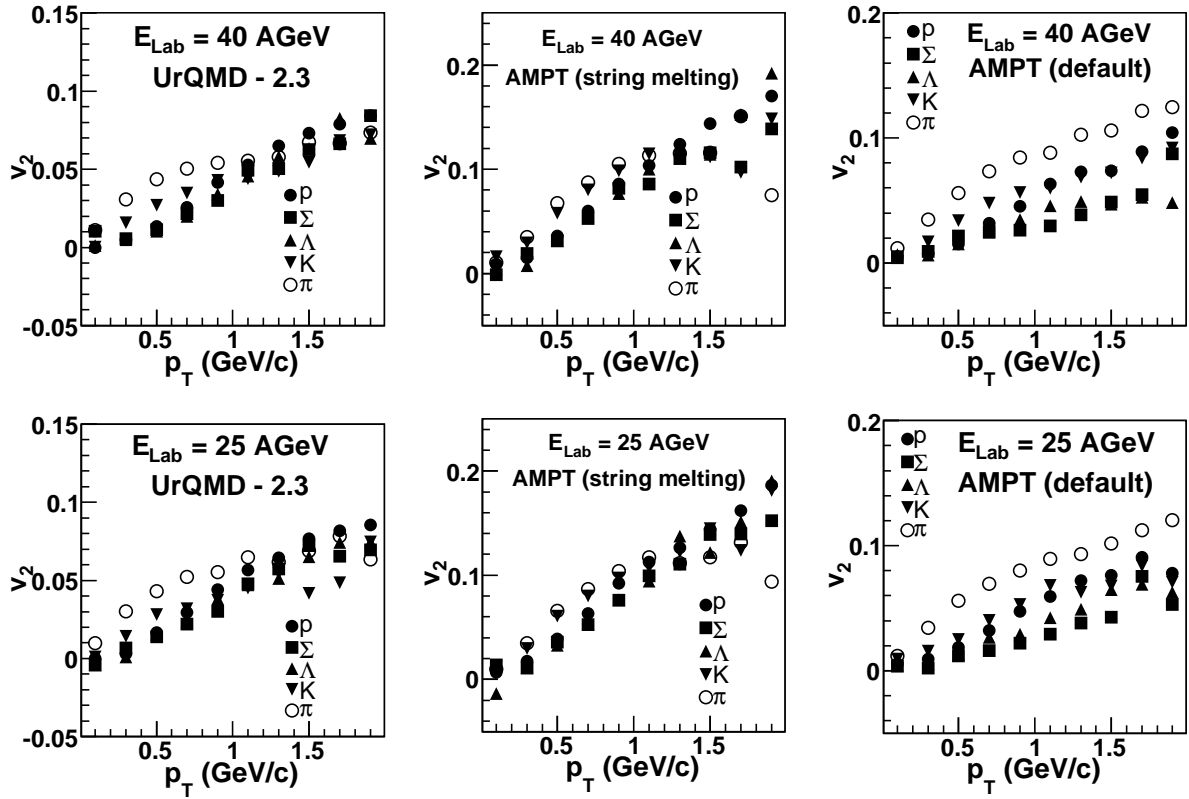}
\caption{\footnotesize variation of $v_2$ with $p_T$ at 40 AGeV (upper panel) and 25 AGeV (lower panel) from different transport models. The figures at the left side show the results from UrQMD whereas figures at middle and right show predictions from string melting and default version of AMPT respectively.}
\label{fig1}
\end{figure*}

In Fig.~\ref{fig1} we present the results for the $p_T$ dependence of $v_2$ of identified hadrons as predicted by UrQMD and AMPT at both top and intermediate FAIR energies. The upper panel shows results at 40 AGeV whereas the lower panel corresponds to 25 AGeV beam energy. In all the cases, we have selected mid-central Au + Au collisions (b = 5 - 9 fm.) and a rapidity window symmetric around centre of mass (CMS) mid-rapidity.

In all the three scenarios, we have found that $v_2$ is an increasing function of $p_T$ at both the energies. In addition, a mass-ordering of $v_2$ for $p_T$ up to $~$ 1.0 GeV/c is observed. At a given value of $p_T$, $v_2$ decreases with increase of hadron mass. In default version of  AMPT without partonic degrees of freedom this ordering is maintained throughout the entire $p_T$ window of investigation. But for UrQMD and AMPT with string melting scenario, the ordering becomes inverse for $p_T> $ 1.2 GeV/c. The heavier baryons are seen to attain larger flow compared to lighter mesons. In this context we would like to mention that, these observations of hadron mass ordering at low $p_T$ and its breaking at high $p_T$ is consistent with the measurements at RHIC~\cite{v2_exp}. Measurement of $v_2$ as a function of transverse momenta $p_T$, showed that at low $p_T$ region, the hadron $v_2$ values exhibit a mass dependent ordering: heavier hadrons have smaller $v_2$. Hydrodynamic model calculations predicted this effect~\cite{v2_hydro}. This mass ordering has been attributed as the signification of a common velocity field. However in the intermediate $p_T$ region (1.5 GeV/c $\le p_T \le$ 5 GeV/c), hydrodynamic results were in gross disagreement with the $v_2(p_T)$ data. The mass ordering appears to switch over, and baryons are seen have larger $v_2$ compared to the mesons; the baryon and meson $v_2$ were found to get separated into two branches. In the FAIR energy range, however the production of high $p_T$ hadrons are small compared to RHIC. We can not go beyond $p_T > $ 2 GeV/c. Even for $p_T > $ 1.5 GeV/c simulated data suffers from large fluctuations.

\begin{figure} \vspace{-0.1truein}
\includegraphics[width=8.6cm]{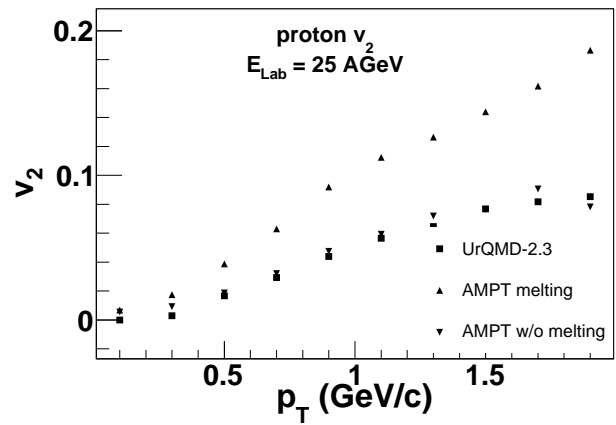}
\caption{\footnotesize variation of $v_2$ of proton with $p_T$ at 25 AGeV obtained from AMPT and UrQMD models. For AMPT both hadronic and partonic versions are used. 
}\label{fig2}
\end{figure}

Among the three cases investigated here, both default AMPT and UrQMD, both of which are based on hadron/string degrees of freedom, give $v_2(p_T)$ very similar in shape and comparable in magnitude. Inclusion of string melting in AMPT which assumes that the initially produced matter is fully partonic in nature results in much higher values of $v_2$ due to a longer phase of partonic interactions in the early reaction phase. For example we have shown in Fig.~\ref{fig2}, the variation of $v_2(p_T)$ for protons at 25 AGeV beam energy. It is clear from the figure that AMPT with string melting scenario, generates $v_2$  values larger compared to default AMPT and UrQMD both of which do not posses partonic degrees of freedom.

We would also like to mention that at top FAIR energy ($E_{Lab} = $40 AGeV), NA49 experiment at CERN-SPS has published their results for $v_2(p_T)$ of pions and protons for Pb+Pb collisions~\cite{NA49}. For completeness we have compared the model predictions, with the available data. The results are shown in the following Fig.~\ref{fig3} .

\begin{figure*} \vspace{-0.1truein}
\includegraphics[width=15.cm]{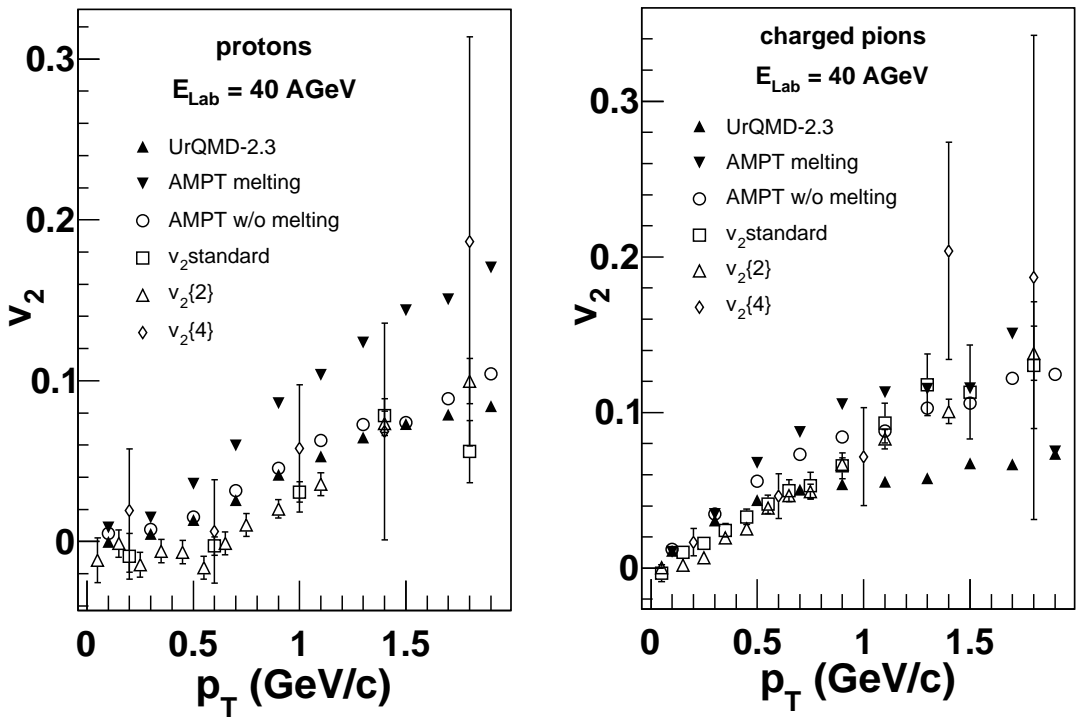}
\caption{\footnotesize variation of $v_2$ with $p_T$ at 40 AGeV. Results from different models are compared with the available data from NA49~\cite{NA49}. The simulated data are for mid-central Au+Au collisions in a rapidity window symmetric about mid-rapidity (-1$\le y_{cm} \le$1) in centre of mass frame. The experimental data however is for mid-central Pb+Pb collisions. The rapidity coverage is slightly different for different methods. The left panel shows results for proton whereas the right panel shows results for charged pions. 
}\label{fig3}
\end{figure*}

The $v_2$ values were calculated from the the data using standard event plane method as well as 2nd and 4th order cumulant methods. The extracted $v_2$ values are found to suffer from large amount of uncertainties especially at high $p_T$. UrQMD is found to underestimate the experimental data at high $p_T$ whereas AMPT with partons is found to slightly overestimate the data. Due to large amount of systematic error in the data it is difficult to draw any conclusive picture. It might be mentioned here that attempts have been made earlier to describe the transverse momentum dependence of elliptic flow at SPS~\cite{urqmd_hydro}. Analysis of the $v_2(p_T)$ data of pions for mid-central (b = 5 - 9 fm.) Pb+Pb collisions at $E_{Lab} = $40 AGeV and $E_{Lab} = $160 AGeV showed that pure transport calculations (UrQMD) under-predict the data especially at high $p_T$; an integrated Boltzmann$+$hydrodynamics approach rather describes the data well.

 But it is clearly visible from the above figures that at both the energies, AMPT with string melting yields larger $v_2$ values compared to UrQMD and default AMPT. This probably signals the enhancement of elliptic flow due to partonic scattering. The build up of $v_2$ is due to different pressure gradients in different directions in the transverse plane. Microscopically the pressure is implemented by the re-scattering among the constituents. For hadronic-string transport models, the constituents are mainly hadrons for which the ingredients (strings and di-quarks) do not interact with others during their formation time~\cite{Zhu_urqmd}. For partonic models the partons interact with each other and with the increase of partonic cross section $v_2$ increases. In AMPT with string melting scenario, the partonic cross section is a free parameter. It has been observed in STAR data analysis~\cite{Voloshin} that the model describes the data well when the cross section is chosen in the range 3 - 10 mb. The best fit is however obtained for a value around 5mb. In our study we have set the partonic cross section as 6mb. It may be noted in this context that UrQMD has been found to under-predict the strength of $v_2$ at RHIC~\cite{Lu_urqmd,Zhu_urqmd}. Analysis of $v_2$ values as a function of centrality showed that this hadron-string transport model fails by 40$\%$ to reach the absolute amplitude of measured $v_2$ at STAR~\cite{star_v2}. The centrality dependencies are however very similar to the data. This underestimation is attributed to the lack of partonic interactions in the model, at hot and dense early stages of the evolution.

\section{Constituent quark number scaling of elliptic flow}

As mentioned earlier, one of the key observations of elliptic flow measurements for Au+Au collisions at RHIC is the approximate quark-number scaling of $v_2$ of identified hadrons~\cite{v2_exp}. Measurement of $v_2$ for identified hadrons as a function of transverse momenta $p_T$ exhibits mass-ordering at low $p_T$ region and a switch over resulting higher $v_2$ values for baryons compared to mesons at the intermediate $p_T$ region, as described in the last section. The baryon and meson $v_2$s are separated into two branches. It is observed that $v_2/n_q$ where $n_q$ is the number of constituent quarks, when plotted as a function of $p_T/n_q$ is approximately same for all identified hadrons. Thus the differential elliptic flow is found to show a remarkable universal scaling with the number of valence quarks.

The observed scaling can be well reproduced with the calculations from the parton coalescence and recombination models for hadronization of the quark gluon plasma~\cite{reco1,reco2,reco3}. The basic assumption of these models is that the invariant spectrum of produced particles is proportional to the product of the invariant spectra of the constituents. It suggests that in the region of $p_T$ where the recombination of the partons dominates the process of hadronization, the effect of mass is minimal, $v_2$ obeys a simple scaling law:
\begin{equation}
v_2(p_T)\approx n_q v_2^{q}(p_T/n_q)
\label{scale}
\end{equation}
where $n_q$ is the number of valence quarks in the hadron and $v_2^q$ is the
elliptic flow parameter for them. 

Such scaling behavior is thus considered as an indication that the early stage when $v_2$ has been build up, is partonic in nature and the effective constituent quarks degrees of freedom play an important role in the hadroniozation process. It has been thus speculated that this scaling is directly related to the formation of color-deconfined phase, where all the hadrons are created at hadronization of the quark-gluon matter, by the recombination or coalescence of partons. In this picture, the quantity $v_2^{q}(p_T/n_q)$ is interpreted as the elliptic flow of the constituent quarks. This means that the transverse expansion of the matter produced in energetic nuclear collisions, develops in a phase dominated by partonic collectivity~\cite{Tian_ampt}. But before drawing a decisive picture, like NCQ scaling indicates color deconfinement in nuclear collisions, one needs to take into account all other possible non-partonic scenarios which may lead to such scaling behavior. At top RHIC energy, an approximate NCQ scaling of the identified hadrons, which describes the data pretty well has already been reproduced by hadronic transport models like RQMD and UrQMD~\cite{Lu_urqmd,Zhu_urqmd}. This observed scaling at top RHIC energy, is attributed to the hadronic cross sections which depend only on the quark content of the colliding hadrons and roughly scale with the number of constituent quarks, as assumed by additive quark model (AQM)~\cite{AQM}.

\begin{figure*} \vspace{-0.1truein}
\includegraphics[width=17.5cm]{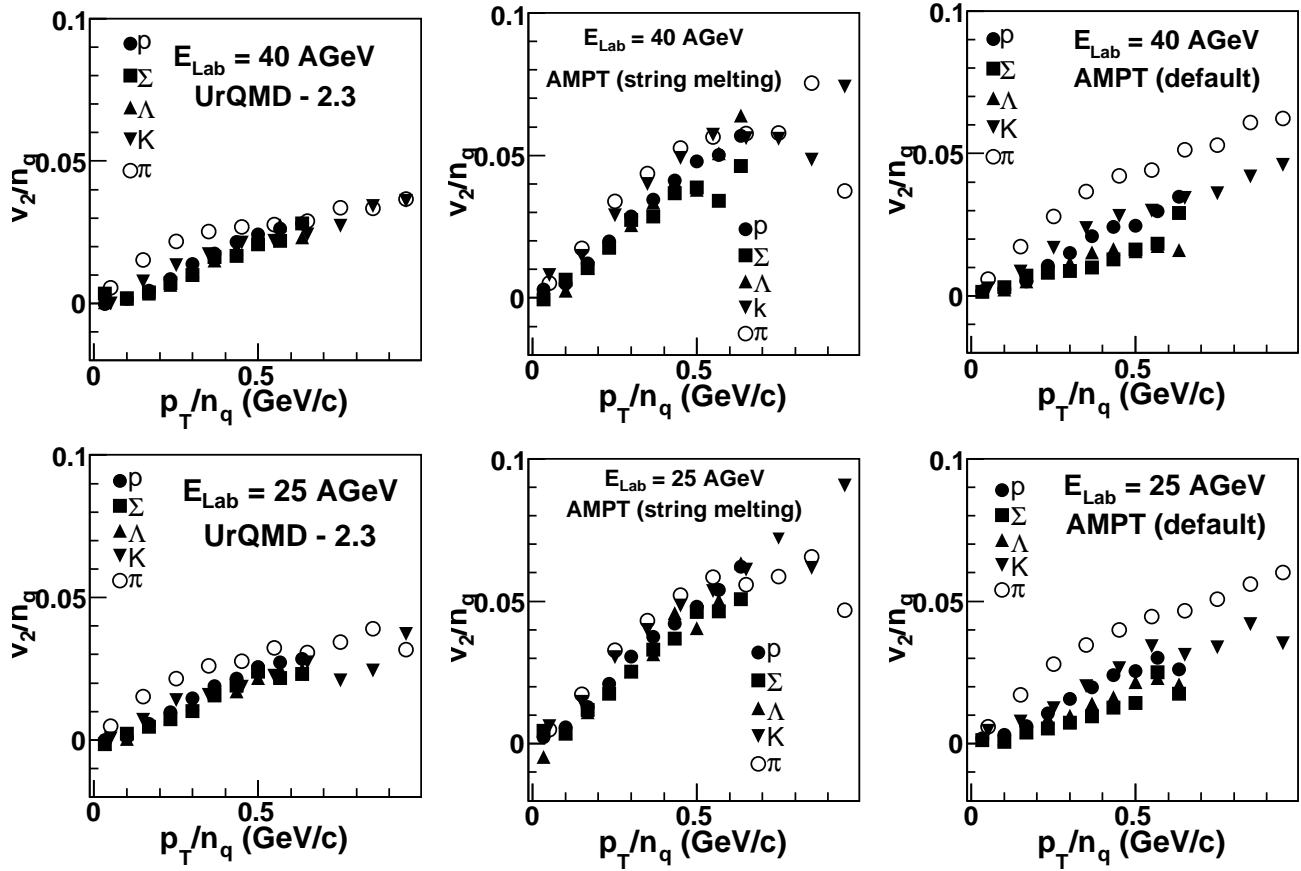}

\caption{\footnotesize variation of $v_2/n_q$ with $p_T/n_q$ at top and intermediate FAIR energies from different models. The upper panel contains results for 40 AGeV and the lower panel contain results for 25 AGeV. In both the cases left figures represent results obtained from UrQMD and middle and right figures show predictions from string melting scenario and default hadronic version of AMPT respectively.}
\label{fig4}
\end{figure*}

Here we have investigated the NCQ scaling behavior as a function of $p_T$  for different hadrons at both the beam energies. In the Fig.~\ref{fig4} we show the variation of $v_2/n_q$ as a function of $p_T/n_q$ at $E_{Lab}$ = 25 AGeV and $E_{Lab}$ = 40 AGeV.

It is observed that over the entire $p_T$ range under study,$v_2/n_q$ as a function of  $p_T/n_q$ does not show any reasonable scaling behavior for either of the models. Results from these models, in the investigated $p_T$ regime, do not show the same behavior that would be expected if quark coalescence would be dominant.

In addition to transverse momentum $p_T$, transverse kinetic energy $KE_T$ is also used at RHIC, as a scaling variable to explore the NCQ scaling behavior of $v_2$~\cite{Phenix_v2_1,Phenix_v2_2}. The transverse kinetic energy, $KE_{T} ( = m_{T} - m_{0})$ is defined as the difference between the transverse mass and the rest mass of a particle, where $m_{T} ( = \sqrt{m_{0}^{2} + p_{T}^{2}}$) is the transverse mass of a particle of rest mass $m_0$. It is considered as a robust scaling variable because it takes care of relativistic effects, which are particularly important for lightest particles. The use of this variable stems out from the idea that the anisotropic pressure gradients in the transverse plane, which give rise to azimuthal anisotropy, lead to collective transverse kinetic energy of the emitted particles. It has been observed at RHIC, that for different particle species, $v_2(KE_T)/n_q$ collapse onto a single curve over the entire range of $KE_T/n_q$. This is in contrast to the $p_T$ scaling which is found to be poor for $p_T/n_q$$\le$ 1 GeV/c and becomes better for $p_T/n_q$$\ge$ 1.3 GeV/c~\cite{Phenix_v2_1}. Even though an excellent scaling has ben observed over the entire range of $KE_T/n_q$ values, this does not imply that quark coalescence describes the behavior from low $p_T$ to intermediate $p_T$ at RHIC. In our current understanding, the origin of $KE_T$ scaling is different and not necessarily related to quark coalescence.

\begin{figure*} \vspace{-0.1truein}
\includegraphics[width=15.2cm]{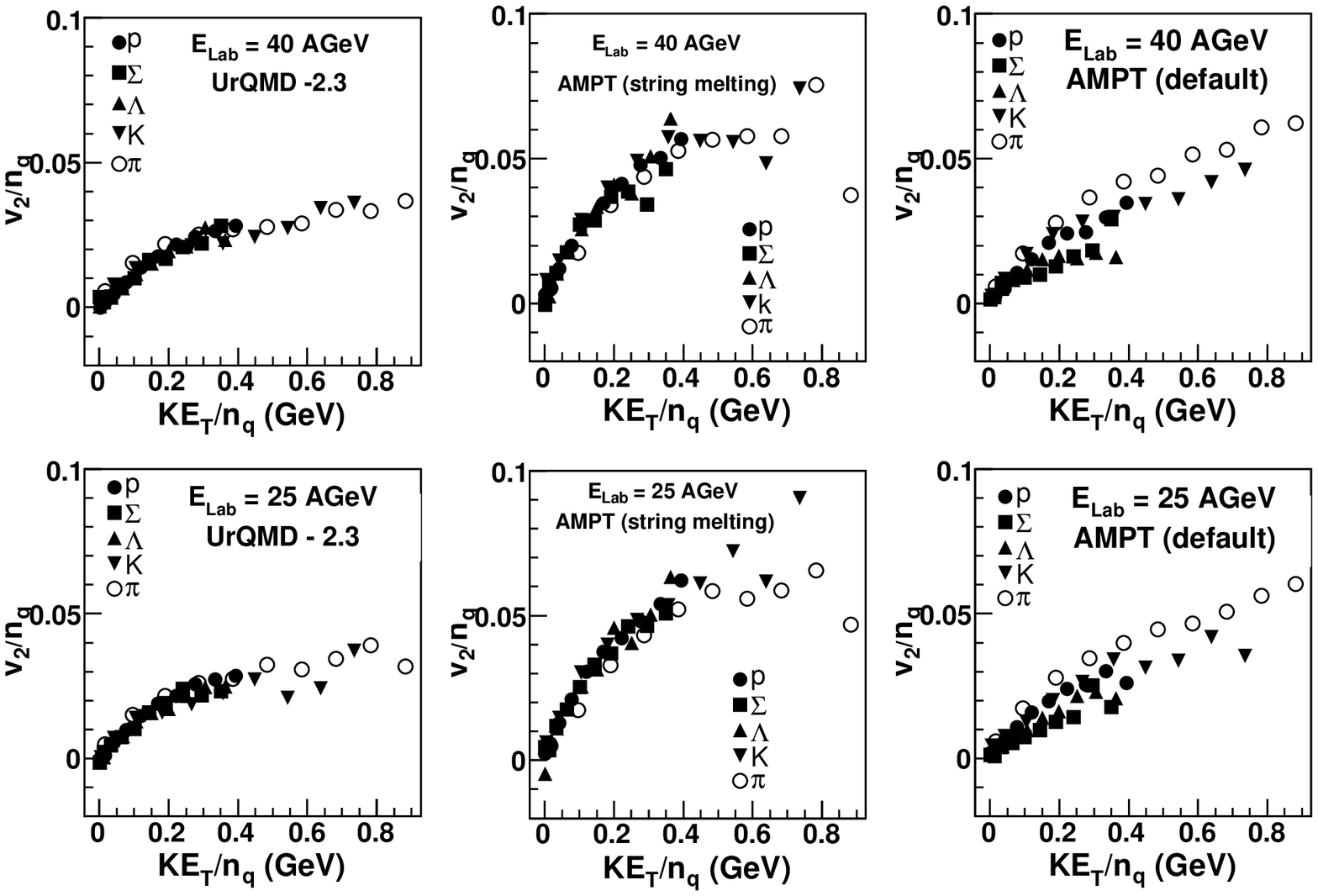}
\caption{\footnotesize variation of $v_2/n_q$ with $KE_T/n_q$ at top and intermediate FAIR energies from different models. The top panel contains results for 40 AGeV and the bottom panel contain results for 25 AGeV. In both the cases left figures represent results obtained from UrQMD and middle and right figures show predictions from string melting scenario and default hadronic version of AMPT respectively. }
\label{fig5}
\end{figure*}

Here we have investigated the NCQ scaling behavior as a function of $KE_T$ for different hadrons at both the beam energies. In the Fig.~\ref{fig5} we show the variation of $v_2/n_q$ as a function of $KE_T/n_q$ at $E_{Lab}$ = 25 AGeV and $E_{Lab}$ = 40 AGeV.

 The default version of AMPT without partonic degrees of freedom does not show any remarkable scaling behavior with respect to $KE_T$ at either of the investigated beam energies. But a reasonable scaling behavior can indeed be produced by both the string melting version of AMPT, as well as UrQMD. Observation of the scaling behavior of $v_2/n_q$ as a function of $KE_T/n_q$ over the entire $p_T$ range under study, is thus in line with the measurements of RHIC, where hydrodynamic mass scaling is believed to be applicable in the low $KE_T$ region  .

UrQMD is a hadronic-string transport model; it does not include quark and gluons as effective degrees of freedom. On the other hand, in AMPT model with string melting scenario, strings are melted into soft partons which can undergo only elastic scatterings among themselves and hence the total number of partons in the system is exactly equal to the number of constituent quarks in the produced hadrons. Thus two different transport models with different degrees of freedom as input are found to exhibit considerable amount of scaling for $v_2/n_q$ of identified hadrons when plotted as a function of $KE_T/n_q$. In view of this observation it appears that the NCQ scaling of hadronic $v_2$ with respect to $KE_T$, will probably not help us to draw any conclusive picture about the nature of the dense nuclear matter expected to be produced in relativistic nuclear collisions at FAIR. Relative values of $v_2$ might play a better role in distinguishing the two scenarios i.e. the partonic scenario and the hadronic scenario.

\begin{figure*} \vspace{-0.1truein}
\includegraphics[width=15.2cm]{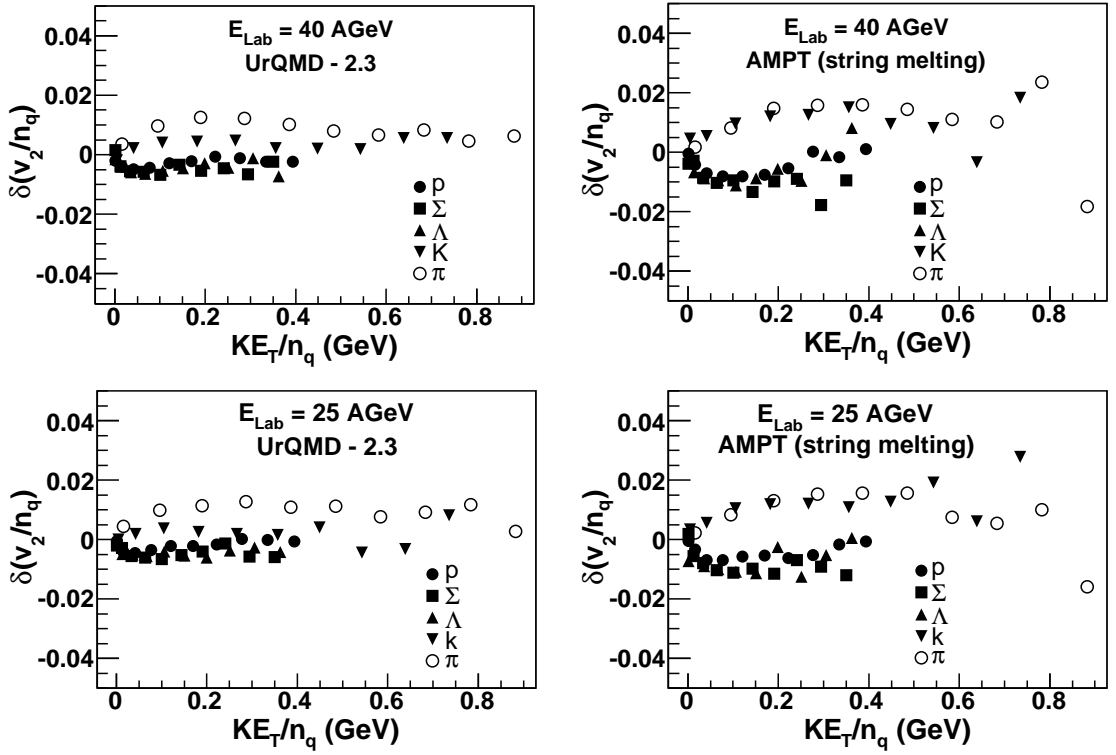}
\caption{\footnotesize variation of $\delta(v_2/n_q)$ with $KE_T/n_q$ at top and intermediate FAIR energies from different models. The top panel contains results for 40 AGeV and the bottom panel contain results for 25 AGeV. In both the cases left figures represent results obtained from UrQMD and right figures show predictions from string melting scenario of AMPT.}
\label{fig6}
\end{figure*}

To extend our study further we have tried to investigate the degree of scaling exhibited by different models. We have plotted the deviation $\delta(v_2/n_q)$, of the scaled $v_2$ values of each hadron from their average value as a function of the scaled $KE_T$. The results for UrQMD and string melting version of AMPT are shown in the Fig.~\ref{fig6}. Since the default version of AMPT does not show any reasonable scaling behavior we have not taken it into consideration. 

For both the models the baryons and mesons are clearly seen to be separated into two branches. Had the scaling been ideal, $\delta(v_2/n_q)$ will be zero for all the particles and they fall on one another. The non-zero value of $\delta(v_2/n_q)$ implies scaling is approximate and it is different for different particle. For the mesons $\delta(v_2/n_q)$ is positive where as for baryons it is negative. However it appears that the deviation is more in case of AMPT compared to UrQMD, which possibly implies that a better scaling is exhibited by UrQMD than AMPT. For a quantitative measure of the degree of scaling, exhibited by these models, we define a variable $\alpha$, the root mean square (RMS) value of the distribution of $\delta(v_2/n_q)$ for all particles over the given $p_T$ range. A better scaling would then result in a lower value of $\alpha$. Table 1 represents the values of $\alpha$ for both the models at the two investigated energies. We have found at both the energies $\alpha_{UrQMD}$ is lower than $\alpha_{AMPT}$ which signifies a better scaling exhibited by UrQMD than AMPT in the FAIR energy regime.

\begin{table}
{\bf Table 1:} Comparison of the values of $\alpha$
                                                                                                    
\begin{tabular}{|c|c|c|} \hline
$E_{Lab}$ (GeV)  & $\alpha_{UrQMD}$ & $\alpha_{AMPT}$ \\
\hline
25 & 0.005633 & 0.01008 \\
\hline
40 & 0.005381 & 0.01025 \\
\hline
\end{tabular}
\end{table}

\section{Summary}

In summary, the differential elliptic flow $v_2(p_T)$ of identified hadrons and its scaling with the number of constituent quarks has been investigated in the FAIR energy regime employing both hadronic as well as partonic transport models.  In both the cases the observations are quite in line with the elliptic flow measurements at RHIC. It has been found that AMPT with string melting which includes partonic scatterings, generates larger flow compared to the hadronic models. None of the models is found to produce the constituent quark number scaling behavior of hadronic $v_2$ with respect to $p_T$, over the $p_T$ range studied in this paper, at either of the investigated energies at FAIR. Both the models exhibit NCQ scaling behavior, at varying degree, with respect to $KE_{T}$, which might be attributed to the hydrodynamics predicted hadron mass ordering of $v_2(p_T)$, valid in the low $p_T$ region~\cite{v2_hydro}. Presence of constituent quark number scaling of hadronic $v_2$ in terms of $KE_T$, in hadronic as well as partonic model makes this observable rather insensitive to distinguish between hadronic and partonic phase at FAIR. A quantitative analysis of the scaling in terms of the scaling variable $\alpha$ further shows that UrQMD exhibits better scaling properties compared to AMPT. If at all a universal scaling behavior of elliptic flow with respect to $KE_T$ is observed at FAIR, whether it should be interpreted as a signature for the formation of a partonic medium or not still remains as a debated issue.


\begin{thebibliography}{99}
\bibitem{sub} Chattopadhyay 2008 J.Phys.G: Nucl. Part. Phys. 35 104027.
\bibitem{density} I. C. Arsene et. al., Phys. Rev. C 75(2007) 034902.
\bibitem{Greiner} H. Stocker and W. Greiner, Phys. Rept. 137, 277 (1986).
\bibitem{Voloshin} S. A. Voloshin, A. M. Poskanzer and R. Snellings, arXiv:0809.2949 [nucl-ex].
\bibitem{Ollitrault} J. Y. Ollitrault, Phys. Rev. D 46, 229 (1992).
\bibitem{v2_sorensen} P. Sorensen, arXiv:0905.0174[nucl-ex]
\bibitem{urqmd} S. A. Bass et al. , Prog. Part. Nucl. Phys. 41 (1998) 255;
  M. Bleicher et. al., J. Phys. G 25, 1859(1999).

\bibitem{ampt} Z. W. Lin and C. M. Ko, Phys. Rev. C65, 034904 (2002);

Z. W. Lin, C. M. Ko, B. A. Lin, B. Zhang and S. Pal, Phys. Rev. C72,064901(2005).

\bibitem{Zhu_urqmd} X Zhu et al., J. Phys.G:Nucl. Part. Phys.32 (2006);
\bibitem{Lu_urqmd} Y. Lu et. al., J.Phys.G32:1121-1130,2006.
\bibitem{AQM} K. Goulianos,Phys. Rep. 101, 169 (1983)

\bibitem{RQMD} H. Sorge, H. Stocker, and W. Greiner, Ann. Phys. \textbf{192}, 266(1989).

\bibitem{QMD} J.Aichelin, Phys. Rep. \textbf{202}, 233 (1991).


\bibitem{Hijing}  {X. N. Wang, M. Gyulassy, Phys. Rev. \textbf{D 44},
3501 (1991).}

\bibitem{ZPC}  {B. Zhang, Comp. Phys. Comm. \textbf{109}, 193.
(1998).}

\bibitem{Pythia}  {T. Sj\"ostrand, Comp. Phys. Comm. \textbf{82}, 74.
(1994).}

\bibitem{ART}  {B. A. Li and C. M. Ko, Phys. Rev. \textbf{C 52}, 2037.
(1995). }
\bibitem{v2_exp}
C.~Adler {\it et al.}, [STAR Collaboration], 
Phys.\ Rev.\ Lett.  {\bf 87} 182301 (2001);
 {\it ibid.} {\bf 89} 132301 (2002);
 \ {\bf 90}, 032301 (2003);

S.~S.~Adler {\it et al.}, [PHENIX Collaboration],
Phys.\ Rev.\ Lett.\  {\bf 91}, 182301 (2003);

S.~Esumi (for the PHENIX collaboration),
Nucl.\ Phys.\  A{\bf 715}, 599 (2003).

\bibitem{v2_hydro} P.F.Kolb, P.Huovinen, U.W.Heinz and H. Heiselberg, Phys.Lett.B 500 (2001)232;
P. Huovinen, arXiv:nucl-th/0305064.

\bibitem{urqmd_hydro} H. Petersen et. al. arXiv:0907.2169 [nucl-th]. 

\bibitem{star_v2}
J. Adams, et al. [STAR Collaboration], Phys. ReV. C72, 014904(2005);
K. H. Ackermann, et al. [STAR Collaboration], Phys. Rev. Lett.86, 402(2001);

\bibitem{Phenix_v2_2} A. Adare, et al. [PHENIX Collaboration], Phys. Rev. Lett.98:162301(2007);

\bibitem{Phenix_v2_1} Michael Issah and Arkadij Taranenko (for the PHENIX Collaboration), Proc. 22nd Winter Workshop on Nuclear Dynamics (2006), arXiv:nucl-ex/0604011;

\bibitem{reco1}
R.~J.~Fries, B.~M\"uller, C. Nonaka and S.~A.~Bass,
Phys.\ Rev.\ Lett. {\bf 90}, 202303 (2003);
V.~Greco, C.~M.~Ko, and P.~L\'evai, Phys.\ Rev.\ Lett.\ {\bf 90},
 202302 (2003);
D.~Molnar and S.~A.~Voloshin,
Phys.\ Rev.\ Lett.\  {\bf 91}, 092301 (2003);
C.~Nonaka, R.~J.~Fries and S.~A.~Bass,
Phys. Lett. {\bf B583}, 73 (2004).

\bibitem{reco2} R.~J.~Fries, B.~M\"uller, C.~Nonaka, and S.~A.~Bass,
Phys. \ Rev. \ C {\bf 68}, 044902 (2003).

\bibitem{reco3} Raktim Abir and Munshi G. Mustafa, Phys. Rev. C 80, 051903 (2009). 

\bibitem{NA49} C. Alt et. al. [NA49 collaboration], Phys. Rev. C 68, 034903 (2003). 

\bibitem{Tian_ampt} J. Tian et. al., Phys. Rev. C 79, 067901 (2009). 

\end{thebibliography}
\end{document}